# On Minimization of a Quadratic Binary Functional

## Leonid B. Litinskii

Center of Optical-Neural Technologies of
Scientific Research Institute for System Analysis of
Russian Academy of Sciences
Moscow, +7(499)135-7802, litin@mail.ru

The problem of minimization of a quadratic functional depending on great number of binary variables is examined. 3 variants of minimization procedure are studied with the aid of computer simulation for spin-glass matrices. It is shown that under other equal conditions evident superiority has *the maximal dynamics* (the greedy algorithm). The dependence of the results on a distance between start points and the ground state is investigated. It is determined that the character of distribution of local minima depends on this distance crucially.

*Keywords:* Energy Landscape, Spin-glass models, Local Search, Global Minimum.

### 1. Introduction

We discuss the problem of minimization of a quadratic form of great number binary variables $s_i$:

$$E(\mathbf{s}) = -\frac{(\mathbf{Js},\mathbf{s})}{2} = -\frac{1}{2}\sum_{i,j=1}^{N} J_{ij} s_i s_j \xrightarrow{\mathbf{s}} \min, \quad \mathbf{J} = (J_{ij})_1^N, \; \mathbf{s} = (s_1, s_2, ..., s_N), \; s_i = \{\pm 1\} . \qquad (1)$$

This problem appears in a lot of fields of science [1]. Finding of maximal cut in graph and resource allocation problem handling the results of physical experiments and many other problems reduce to minimization of the functional (1). In physics of magnetic materials [2] this problem appears with regard to finding the ground state that is such a binary vector **s** that provides the global minimum of the functional (1).

Without restriction of generality the connection matrix $\mathbf{J} = (J_{ij})_1^N$ can be regarded as a symmetric one with zero elements of the main diagonal: $J_{ij} = J_{ji}$, $J_{ii} = 0$. Using physical terminology the binary variables $s_i = \{\pm 1\}$ will be called *spins*, $N$-dimensional vectors $\mathbf{s} = (s_1, s_2, ..., s_N)$, which define the state of the system as a whole, will be called configuration vectors or *configurations*, the functional $E(\mathbf{s})$ will be called *the energy* of the state **s**, and the configuration providing the global minimum of the functional (1) will be called *the ground state*.

Each problem generates its own specific connection matrix **J**. The common for all these matrices is the absence of an effective computational procedure which allows one to find the ground state during a reasonable time. The problem is that usually the number of local minima is exponentially large in *N*, and it is very difficult to get the deepest minimum. An exhaustive search is possible only for small dimensions $N \sim 20$. For large dimensions $N \sim 10^2 - 10^3$ one restrict himself by finding the deepest local minimum being not guaranteed that the obtained minimum is the global one. A large number of papers are devoted to working out effective methods of minimization of the functional (1). With regard to physical problems to get the idea of the latest progress in this filed one can use the recent monograph [3]. The revue of previous works can be found in [4]-[6].

Perhaps, the greatest progress is achieved for the Edwards-Anderson spin-glass model (the EA-model) [7]. In this model spins are placed in vertices of a regular lattice (usually it is a quadratic or a cubic ones). Besides that it is supposed that only the nearest spins are interacting. The most elements of the connection matrix are equal to zero, and nonzero elements are chosen randomly from the standard normal distribution. For such matrices the effective algorithm is developed [8], which allows one even for very large dimensionalities (*N* > 2000) to find the ground state during a reasonable time. There is the internet-website where one can send his matrix, and as an answer he will get the ground state of the functional (1) (see [9]). We used the facilities of this website.

The other interesting object investigated by physicists is the Sherrington-Kirkpatrick model of the spin glass (the SK-model) [10]. Under this approach it is assumed that every spin interacts with all other spins, and the elements of the connection matrix are random numbers that are taken from the standard normal distribution. In this case the matrix is filled up densely. From the physical point of view this model is not a realistic one, however its properties are investigated theoretically. The energy surface is the system of hierarchically embedded "valleys", and the number of local minima increases exponentially when *N* increases. About the ground state of the SK-model we only know that it has to be a non-degenerate one. Neither the corresponding configuration nor the energy of global minimum is known. In this problem there is no algorithm allowing us to find the ground state for large dimensionality *N*. When it is necessary to know the global minimum, a large number of random starts are done, and the configuration corresponding to the deepest minimum is taken as the ground state (see, for example, [11], [12]). In our experiments we used the same approach (see Section 4).

Concluding this short list of modern minimization methods, let us point out very interesting the GRA-algorithm, which is developed by the authors of [13], [14]. In spite of the fact that this method is a heuristic one, it is based on fundamental physical ideas and as it follows from publications shows good results. It is important that the GRA-algorithm is universal and does not depend on the type of the connection matrix.

Any minimization algorithm is based on either one or another variant of the procedure of the local minimization. That is the system starts from an initial state $\mathbf{s}(0)$, then we flip all spins one after another and check if the value of the functional (1) decreases or not. If it is decreases, this spin assigns the new (flipped) value. If the value of the functional does not decrease, the previous value of the spin variable is restored. The procedure of the local minimization can be organized in different ways. The result of minimization depends on the way of its realization [13]-[18]. The main goal of our paper is the experimental analysis of three variants of the procedure of local minimization. In the next Section we formalize the problem. In other Sections we describe and discuss the results of computer simulations.

In the papers [19], [20] the energy landscape of the functional (1) was analyzed theoretically for the Hopfield model of the neural network. The authors used the probability-theoretical approach, which is used for estimation of the storage capacity of the Hopfield model [1]. They succeeded in obtaining a lot of interesting results about minima of the functional (1). In particular, it turned out that it is possible to connect the depth of the local minimum with the probability of its random finding. The obtained analytical expression indicates that the deeper the local minimum, the wider the basin of its attraction and, consequently, the larger the probability of its random finding. In spite of seeming simplicity of this statement, it is rather nontrivial. For example, from it follows that under a great number of random starts the probability to get into the ground state (or near it) is greater than the probability to get into less deep minima. Up to the end of the paper this statement will be called *the rule of deep minima*.

For the Hebb connection matrices, used in the Hopfield model, the authors of [19], [20] obtained a good agreement of their theory with computer simulations. In the present time the authors examine other types of connection matrices [21]. In our paper we verify when the rule of deep minima is true for spin-glass matrices. This is an additional goal of our work.

## 2. Different dynamics

For the first time the procedure of the local minimization of the functional (1) was used in [22]. In the theory of neural networks this procedure is known as the consequent dynamics. One starts the system from the initial configuration $\mathbf{s}(0)$, in random order examine the spins, and consequently gives to each spin the value that is equal to the sign of the local field $h_i(t) = \sum_{j=1}^{N} J_{ij} s_j(t)$. Formally the dynamical rule has the form

$$s_i(t+1) = \begin{cases} s_i(t), & \text{when } s_i(t)h_i(t) \geq 0; \\ -s_i(t), & \text{when } s_i(t)h_i(t) < 0. \end{cases}$$

If the sign of the spin does not coincide with the sign of the field acting onto this spin, if the spin is *dissatisfied*, $s_i(t)h_i(t) < 0$, one flips it $s_i(t+1) = -s_i(t)$. Otherwise the value of the spin does not change and we pass on to the next spin. This algorithm is one of algorithms investigated in this paper. It will be called *the random dynamics*.

It is easy to see that when we flip a dissatisfied spin the energy of the state decreases:
$$E(\mathbf{s}(t+1)) = E(\mathbf{s}(t)) - 2|h_i(t)|. \tag{2}$$
Sooner or later the system finds itself in the state where all spins are satisfied and evolution of the system stops. This state is the local minimum of the energy.

*The maximal dynamics,* which is also examined in this paper, differs from the random dynamics. In this case at every step we flip only those dissatisfied spin which leads to maximal decrease of the energy. In other words, if in the state $\mathbf{s}(t)$ there are some dissatisfied spins, we flip only those to which the maximal modulus of the local field $h_i(t)$ corresponds (see Eq.(2)). The rationale to introduce the maximal dynamics is obvious: If at every step we decrease the energy $E(\mathbf{s})$, and our goal is to find the deepest minimum, *the maximal decreasing of the energy at every step* can be successful. The maximal dynamics is known as greedy algorithm [15], [16].

At last we examined one more variant of the dynamics, *the minimal dynamics*, when we flip those dissatisfied spin, which leads to minimal decrease of the energy [16]. In the paper [16], where such dynamics was called the reluctant-algorithm, it was established that for the SK-model this dynamics gave the best results comparing with other dynamics. This result was a surprise for the authors of [16] themselves, so it was important to verify it [17], [18]. Looking forward, let us say that really for the SK-model the minimal dynamics gives the best results, only *if we start the dynamical system from the point of the general position*. If we move start points nearer to the ground state, the minimal dynamics loses comparing both random and maximal dynamics.

Over all the paper we will use this three introduced above terms: random, maximal and minimal dynamics. We compared these dynamics minimizing the functional (1) for matrices corresponding to the two dimensional EA-model with periodical boundary conditions and for the SK-model (see Introduction). Though both types of the matrices are modeling a spin-glass, we keep the name *spin-glass matrices* only for the matrices of the EA-model. For convenience, matrices relating to the SK-model will be called the *Gaussian matrices*.

We tried to find out how the result of minimization depends on the type of the matrix and its dimensionality $N$: Which of dynamics and under which conditions give the best result? Previously, for each matrix we found the ground state of the functional (1). This allowed us to set a question: How the result of minimization depends on the distances of the start points from the ground state? Always (apart of specially specified cases) we measure the distance between two configurations **s** and **s'** by *the relative Hamming distance*:

$$d = \frac{N - abs(\mathbf{s}, \mathbf{s}')}{2N}, \quad \text{where } (\mathbf{s}, \mathbf{s}') = \sum_{i=1}^{N} s_i s_i'. \tag{3}$$

In other words, $d$ is a relative number of coordinates, which differs for two different configurations, $d \in [0, 0.5]$. Configurations, which are at the distant 0.5 from each other, are orthogonal.

All three abovementioned dynamics belong to one-step procedures, when at every step only one spin is flipped. Sometimes $k$-optimal procedures are examined. In this case groups of $k$ spins are flipped as a whole, or even procedures with variable depth of search, when the size of flipped group is not fixed, but is chosen dynamically [13], [14]. We examine one-step procedures only.

### 3. Results for spin-glass matrices

In this section the problem in question is random matrices relating to the two-dimensional EA-model of spin-glass with periodical boundary conditions and nearest neighbor interaction. We examined 9 matrices in threes matrices of dimensionality $N$ = 100, 400 and 900. Due to possibilities granted by the web-site [9], for every matrix we knew the ground state of the functional.

For each matrix we generated 10000 random start configurations outlying at the distance $d$ from the ground state and started all three dynamics: the random, the maximal and the minimal ones. For each of dynamics we obtained 10000 local minima for which we recorded the energies $E$, the distances $D$ to the ground state and the number of steps $S$ after which the system occurs in the local minimum (see below why the characteristic $S$ is interesting). All these were done for *start distances*: $d$ = 0.05, 0.1, 0.2, 0.3, 0.4 and 0.5; always by start distance we meant the distance from the start configuration to the ground state.

**1)** The detailed results for one of the matrices of the dimensionality $N$=100 are given in Fig. 1-6. Let us explain what is shown on the different panels of the figures. The left columns of each figure correspond to the maximal dynamics, the middle ones correspond to the random dynamics and the right columns correspond to the minimal dynamics. Everywhere along the abscissa axes we show the energy of local minima $E$. For this matrix the ground state energy is equal to $E_{GS} = -1.232$. On the upper panels along the axes of ordinates we put the probability to get into a local minimum. (More correctly, it is the frequency to get into a local minimum, calculated with the aid of 10000 events.) Here we also show the number of different local minima $K_{lm}$ and the energy of the deepest one $E_{min}$. On the lower panels along the axes of ordinates we show the distance $D$ from a local minimum to the ground state. Thus on the lower panels of the figure one can see the distribution of local minima on the coordinate plane $\{E, D\}$.

**1.a)** In Fig.1 the results for starting from the distance $d$=0.05 are shown. As we see, the ground state can be found for all three dynamics ( $E_{min} = E_{GS}$ for each of the dynamics). However, the probability to get into ground state for the maximal dynamics (~0.9) is 3 times greater then the analogous probability for the random dynamics (~0.3), which is 2 times greater than the probability to get into the ground state for the minimal dynamics.

For the maximal dynamics we got only 43 local minima. It is 20 times less than the number of minima, which were found for the random dynamics. The last is two times less than the number of minima found for the minimal dynamics. Abovementioned 43 minima are the deepest ones. They are localized at the left side of a wide energy interval that is filled up with the energies of local minima obtained for random and/or minimal dynamics. It can be said that for the start distance $d$=0.05 on average the maximal dynamics allows us to get deeper minima than the random or the minimal dynamics.

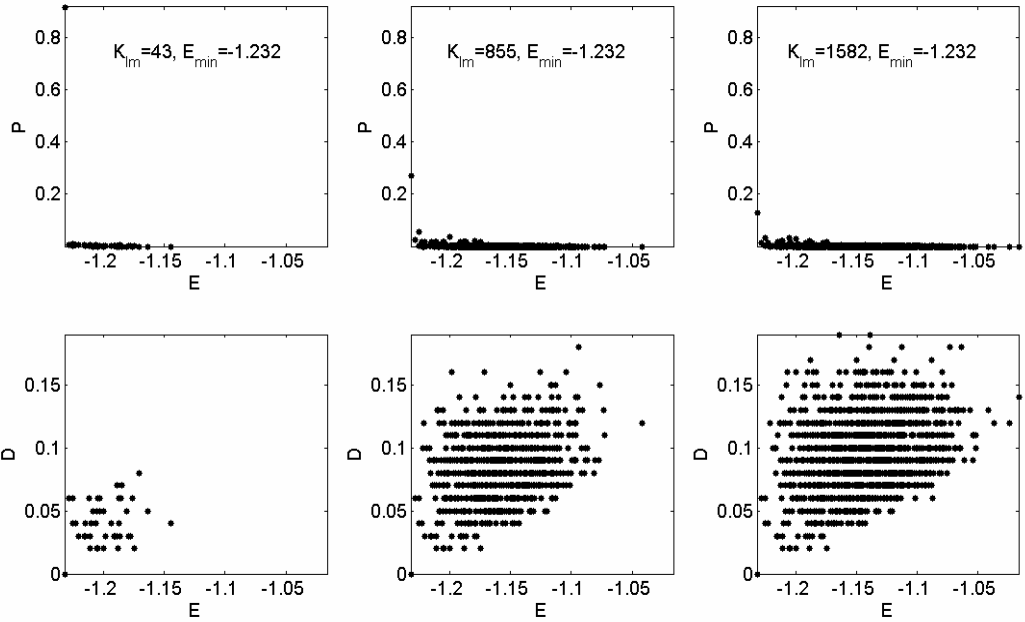

Fig.1. The results of 10000 starts with the distance $d$=0.05 from the ground state for spin-glass matrix ($N$=100). At the left panels we show the results for the maximal dynamics, at the centre panels for the random dynamics and at the right panels for the minimal dynamics. Energies of local minima $E$ are along the abscissa axis. Along the ordinate axes we have: at the upper panels the probability $P$ to get into a local minimum; at the lower panels the distance $D$ from the local minimum to the ground state. $K_{lm}$ is the number of different local minima, $E_{min}$ is the minimal energy we found.

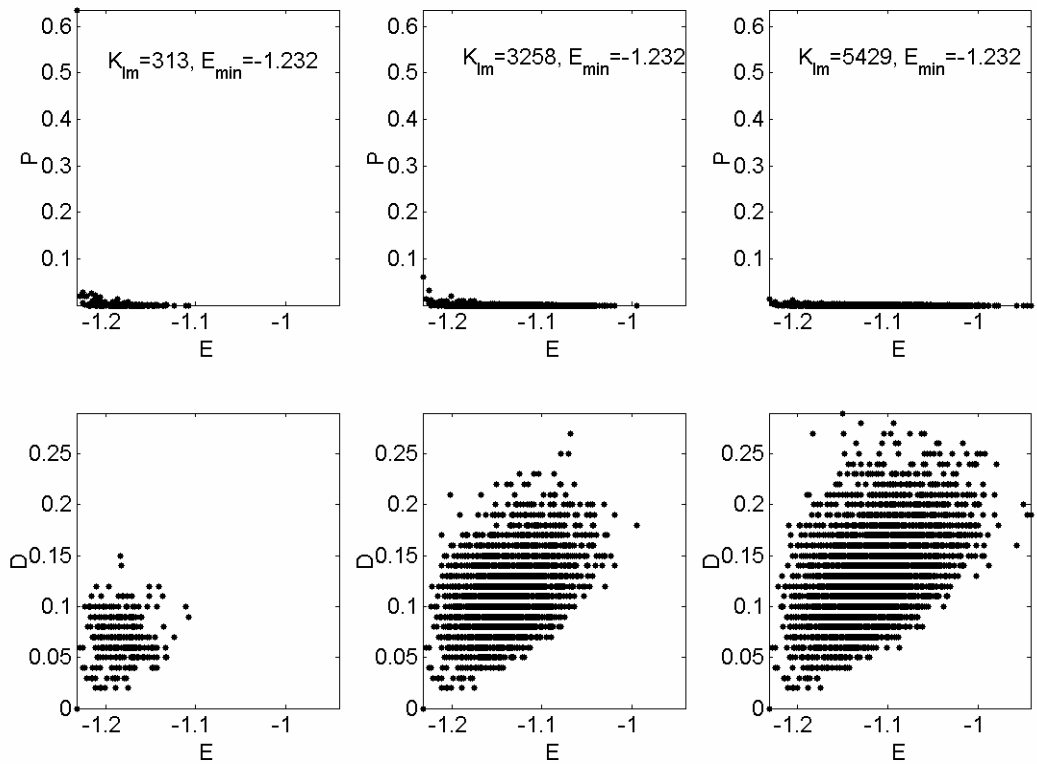

Fig.2. The same as in Fig.1 for the start distance $d$=0.1.

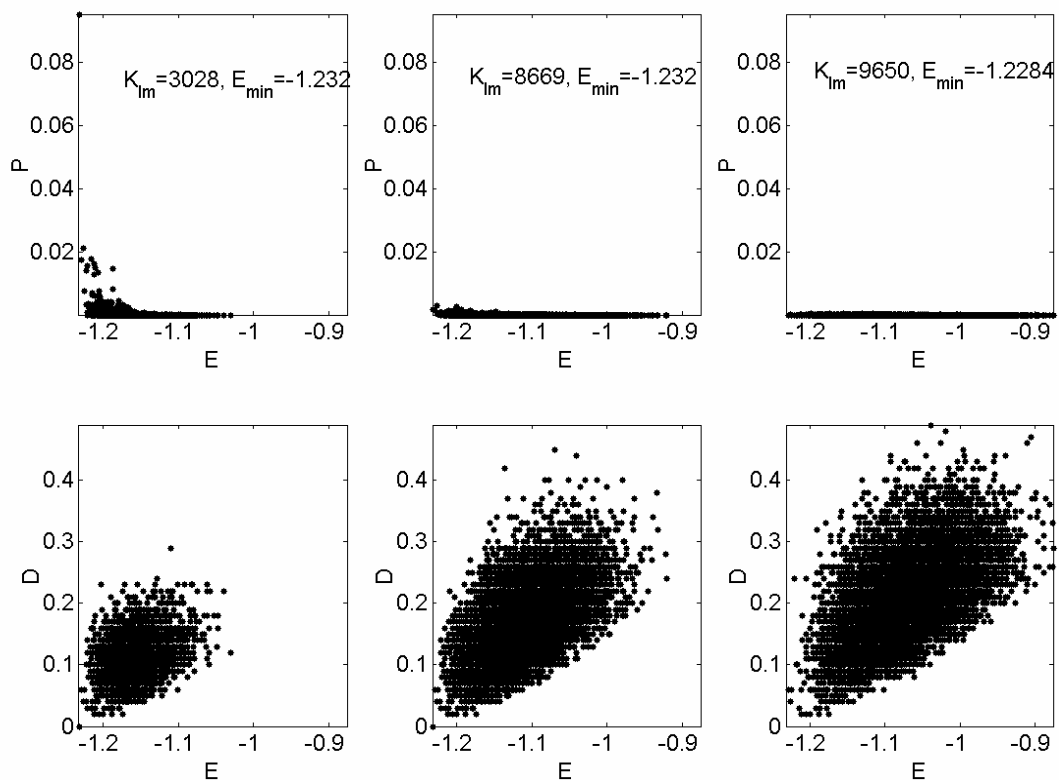

Fig.3. The same as in Fig.1 for the start distance *d*=0.2.

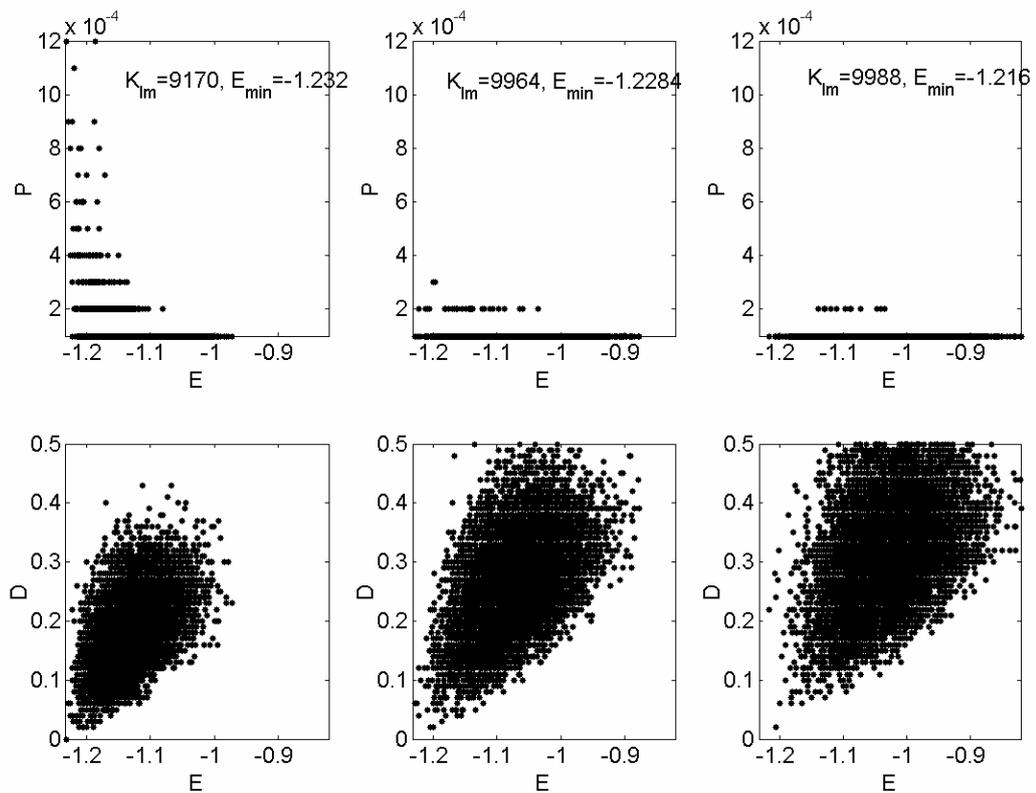

Fig.4. The same as in Fig.1 for the start distance *d*=0.3.

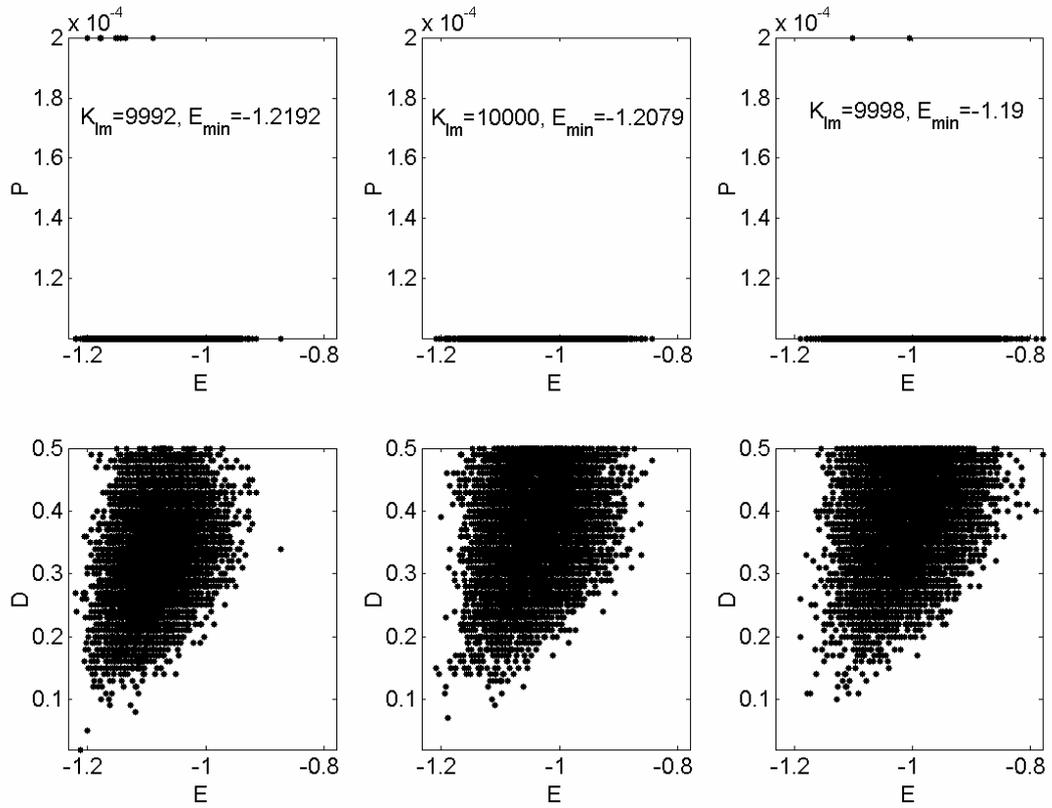

Fig.5. The same as in Fig.1 for the start distance $d$=0.4.

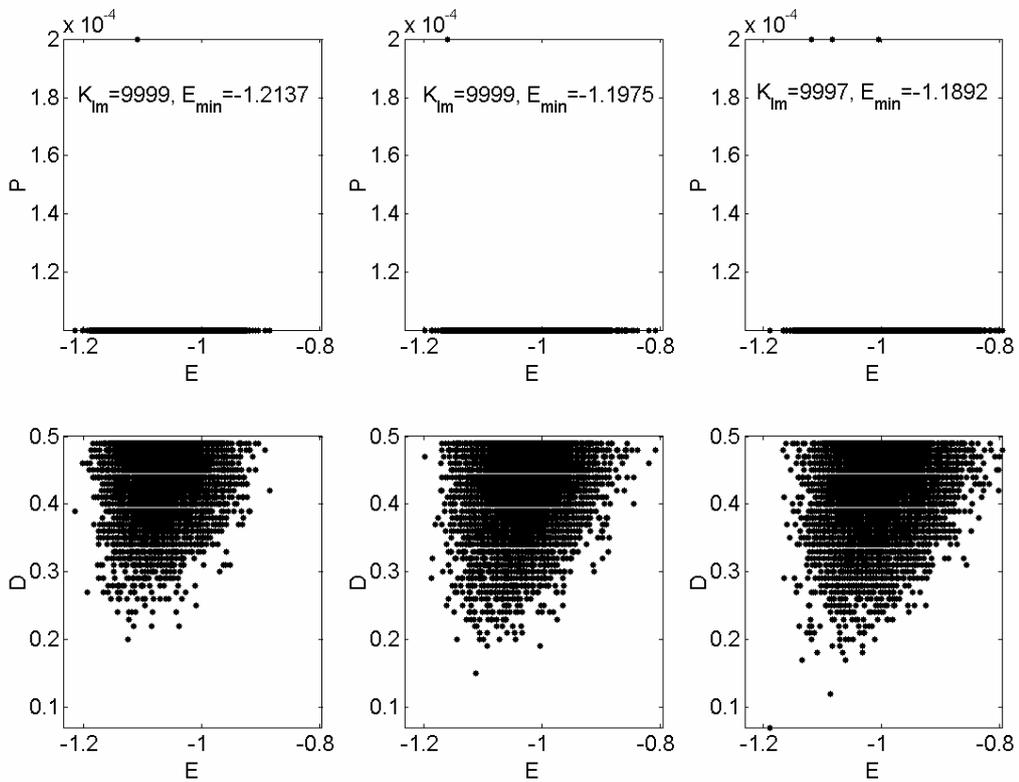

Fig.6. The same as in Fig.1 for the maximal distance $d$=0.5.

Let us analyze the lower panels of Fig.1. We can see that the most part of the 43 minima, obtained with the aid of maximal dynamics are closer to the ground state than start points (for which $d=0.05$). On the contrary, the most part of minima that have been found with the aid of random and minimal dynamics are farther from the ground state than start points. As a rule for these local minima the inequality is true: $D > 0.05$. It can be said that the maximal dynamics shifts the system toward the ground state, while in the case of the random (or the minimal) dynamics, as a rule the system moves away from the ground state.

All aforesaid can be repeated for the starts from the distances $d=0.1$ and $d=0.2$ (see Fig.2 and Fig.3). As the start points move away from the ground state the superiority of the maximal dynamics little by little decreases. However, it can be said that for start distances $d \leq 0.2$ the maximal dynamics exceeds both the random and the minimal dynamics substantially.

For the start distance $d=0.3$ the maximal dynamics still exceeds all other dynamics noticeably (see Fig.4). Indeed, for this $d$ one succeeds in finding the ground state only with the aid of maximal dynamics. Moreover, for the maximal dynamics the distribution of local minima at the coordinate plane $\{E, D\}$ is more compact than for other dynamics (compare the lower panels in Fig.4). However, let us point out that for the maximal dynamics the number of local minima $K_{lm}$ also becomes practically equal to the number of starts $K$: $K_{lm} \approx K = 10000$.

For even larger start distances, $d=0.4$ и $d=0.5$, the difference between these three dynamics ceases to be noticeable (see Fig.5 and Fig.6). Though the deepest local minima are found with the aid of maximal dynamics, total numbers of local minima and the character of their distribution at the plane $\{E, D\}$ are practically the same for all three dynamics.

**1.b)** Let us note that for start distances $d \leq 0.3$ the probability to get into most deep local minima is larger than the probability to get into less deep minima. A characteristic peak in the beginning of the graphs at the upper panels of Figs. 1-4 indicates this. In other words, for these start distances the rule of deep minima is fulfilled (see two last paragraphs of Introduction). On the contrary, for start distances $d > 0.3$ the rule of deep minima does not fulfilled. The graphs on the upper panels of Fig.5 and Fig.6 indicate this. Here the number of local minima practically does not differ from the number of starts $K=10000$. Now all minima become equiprobable: The probability to get into every minimum is $\sim 1/K$.

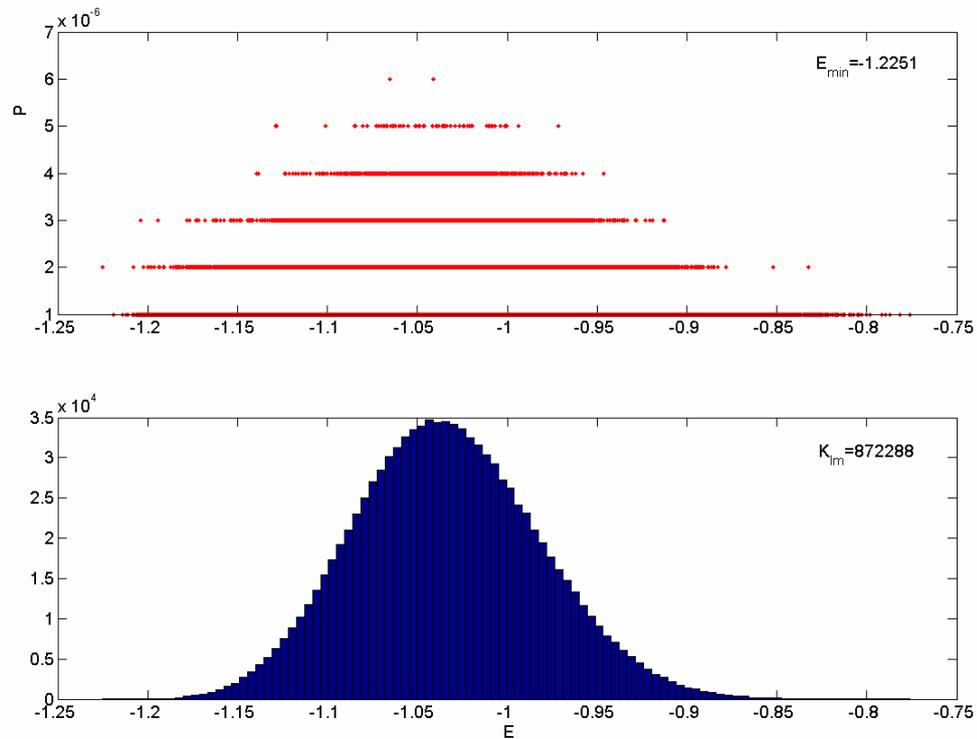

Fig.7. The results of $10^6$ starts for the spin-glass matrix ($N=100$). At the lower panel we show the histogram of the density of distribution of local minima as function of energy; at the upper panel are probabilities to get in local minima.

May be the failure of the rule of deep minima is related to the poor statistics? May be 10000 starts is insufficient? To clarify the situation an extra experiment has been done. Namely, for the distance *d*=0.5 the random dynamics was started 1 million times. The results of this experiment are shown in Fig.7. At both panels along abscissa axes we show the energy of the local minima. On the lower panel the density histogram of the local minima energies is presented. Here we also present the total number of local minima $K_{lm}$. On the upper panel along the ordinate axis is the probability to get into local minima. Here we also write down the energy of the deepest minimum we found. (All calculations were performed using 8 nonzero decimal digits of mantissa.)

We see that with the aid of 1 million starts we got enormous number of local minima: $K_{lm} \approx 900000$, but the ground state was never found ($E_{\min} = -1.2251 > E_{GS} = -1.232$). The graph at the upper panel of the figure testifies that the probability to get in the deepest local minima is less than the probability to get in a minimum that is in the middle of the energy interval. In other words, here the rule of deep minima does not fulfill.

<u>Summarizing the aforesaid.</u> For start distances *d* that are less than a critical value $d_c$, the rule of deep minima fulfilled. In this case the maximal dynamics notably surpasses other two dynamics. On the contrary, for start distances that are larger then the critical distance, $d > d_c$, the rule of deep minima does not fulfill, and the probability to get in the deepest minima is very small. Here the superiority of maximal dynamics becomes hardly noticeable. Approximate equality of the number of local minima and the number of the starts, $K_{lm} \approx K$, indicates that we are in the region $d > d_c$.

The described picture is also reproduced for other two matrices of dimensionality *N*=100: when the start distance does not exceed the critical value $d_c \approx 0.3$, the maximal dynamics surpasses the random dynamics as well as the minimal dynamics. When the start distance increases the superiority of the maximal dynamics little by little decreases, and in the region $d > d_c$ the difference between the dynamics practically vanishes.

**2)** In what follows we will use more compact method of showing our results. We will place at one graph the results for all start distances $0.05 \le d \le 0.5$. We will concentrate at the analysis of some basic characteristics. In Fig.8 for all three dynamics the dependence of these characteristics on the start distance *d* is shown. Not to duplicate what has been said above, in the figure the results for the second spin-glass matrix of dimensionality *N*=100 are shown.

In the figure the solid line corresponds to the maximal dynamics, the dashed line corresponds to the random dynamics and the dash-dotted line corresponds to the minimal dynamics. At two upper panels and the left lower panel along the abscissa axes we show the start distance *d* in terms of the relative Hamming distance (3). At the lower right panel we used the ordinary Hamming distance $d \cdot N$. It is equal to the number of different coordinates of two configurations. The reason why we did exception for this panel would be clear below.

At the left upper panel along the axis of ordinates we show the probability *P* to find the ground state. It is seen that at first the maximal dynamics dominates over other dynamics. Then, when the start distance becomes more than critical distance $d_c$ all three dynamics become equal.

At the lower left panel along the axis of ordinates we show the number of different minima found for each of dynamics. It is seen that the random dynamics (the dashed line) and the minimal dynamics (the dash-dotted line) much quicker than the maximal dynamics come out to the regime where the number of local minima ceases to differ from the number of starts: $K_{lm} \approx K = 10000$. For *d* > 0.3 this equality is also true for the maximal dynamics.

At the right upper panel along the axis of ordinates we show the difference between the energy of the ground state $E_{GS} = -1.2788$ and the minimal energy found after 10000 starts: $E_{GS} - E_{\min}$. For small start distances *d* (when the ground state can be found using 10000 starts) this difference is equal to zero for all dynamics (see initial parts of the graphs). If the start distance increases and 10000 starts are not enough to find the ground state, the difference between energies becomes negative. However, the curve relating to the maximal dynamics dominates other curves. In other words, here the maximal dynamics demonstrates better results too, since it allows us to find deeper minima.

At last at the lower right panel along the axis of ordinates we show the averaged number of steps <*S*> after which the system finds itself in the ground state. In fact we analyze the length of a trajectory that leads from the start point to the ground state. This length is measured by the number of flipped spins (the number of steps). Of course, the curves at this graph are defined up to the start distances that still allow us to find the ground state. For the minimal dynamics this corresponds to $d \cdot N = 20$, for the random and the maximal dynamics we have $d \cdot N = 30$ (compare with graphs at upper right panel). Let us explain why the characteristic *S* is interesting.

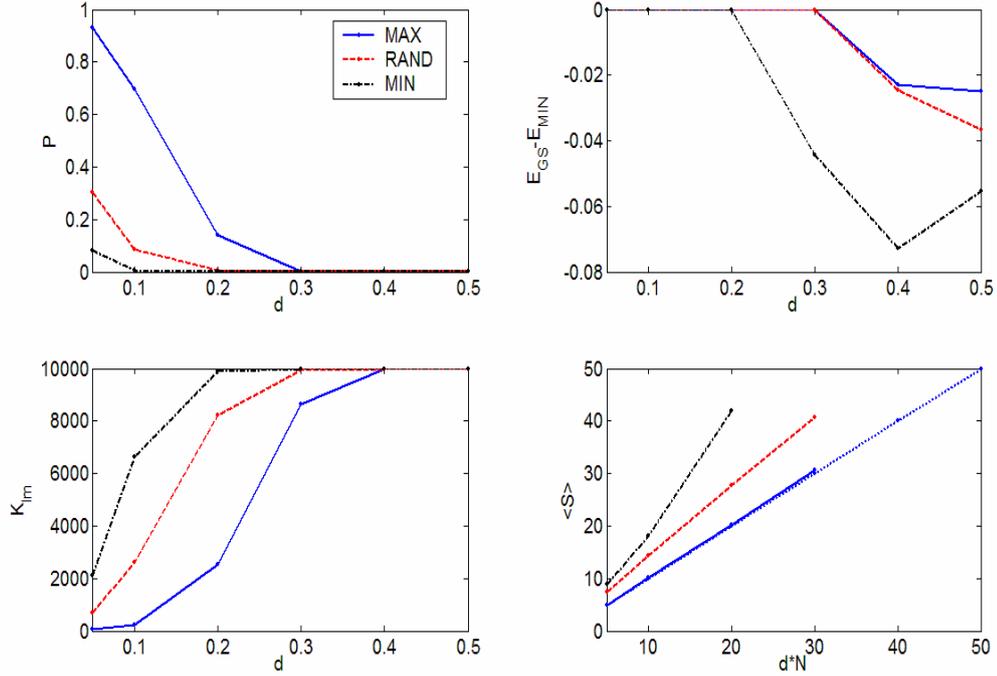

Fig.8. For the spin-glass matrix ($N=100$) the dependence of the results of minimization as functions of the start distance from the ground state. Along the abscissa axis we have: at the right lower panel the Hamming distance $dN$, at other panels the relative Hamming distance $d$ (3). Solid lines correspond to the maximal dynamics, dashed lines correspond to the random dynamics, and dash-dotted lines correspond to the minimal dynamics (see explanation in the body of the paper).

Since we know the start distance $d \cdot N$, it is interesting to understand along which trajectory a descent to the ground state occurs. One thing, if the dynamic system moves to the ground state directly, reducing the distance to it at every step of evolution. The length of such a trajectory approximately is equal to the start distance $S \approx d \cdot N$. It's quite another matter, if the system moves along more complicated trajectory sometimes approaching the ground state, sometimes moves away from it. The movement can be even a chaotic one, only little by little approaching the system to the ground state. Then the number of spins flipped during the descent would be greater than the start distance $S > d \cdot N$. At the right lower panel of Fig.8 just the averaged length of the trajectory is shown. This is the trajectory along which the dynamic system gets to the ground state. We averaged over all hits in the ground state. Point line that is going to the right upper corner of the graph is the bisector $<S> = d \cdot N$. It is convenient to compare all the curves with this line.

We see that the $<S>$-curve for the maximal dynamics is practically the same as the straight line $<S> = d \cdot N$. In other words, when the maximal dynamics is used, the system moves along the shortest trajectory just into the ground state. In the case of the random dynamics analogous characteristic is larger than the start distant $d \cdot N$. And for the minimal dynamics it is even greater. However, as a whole for all three types of dynamics the values of these characteristics differ not significantly. Another situation we have for Gaussian matrices (see the next Section).

Just the same picture has a place for spin-glass matrices of larger dimensionality. In Fig.9 the analogous graphs for one of matrices of dimensionality $N=400$ are shown, and in Fig.10 for one of matrices of dimensionality $N=900$. On the graphs one can see all the characteristic features noted above for matrices of dimensionality $N=100$. Only the critical value of the start distance $d_c$, beginning from which the number of local minima $K_{lm}$ becomes equal to the number of starts $K$, changes. According the results of our calculations where all 3 matrices were used, the critical distances can be estimated as $d_c \approx 0.2$ for $N=400$, and $d_c \approx 0.15$ for $N=900$. These results can be done accurately in more detailed computer simulations.

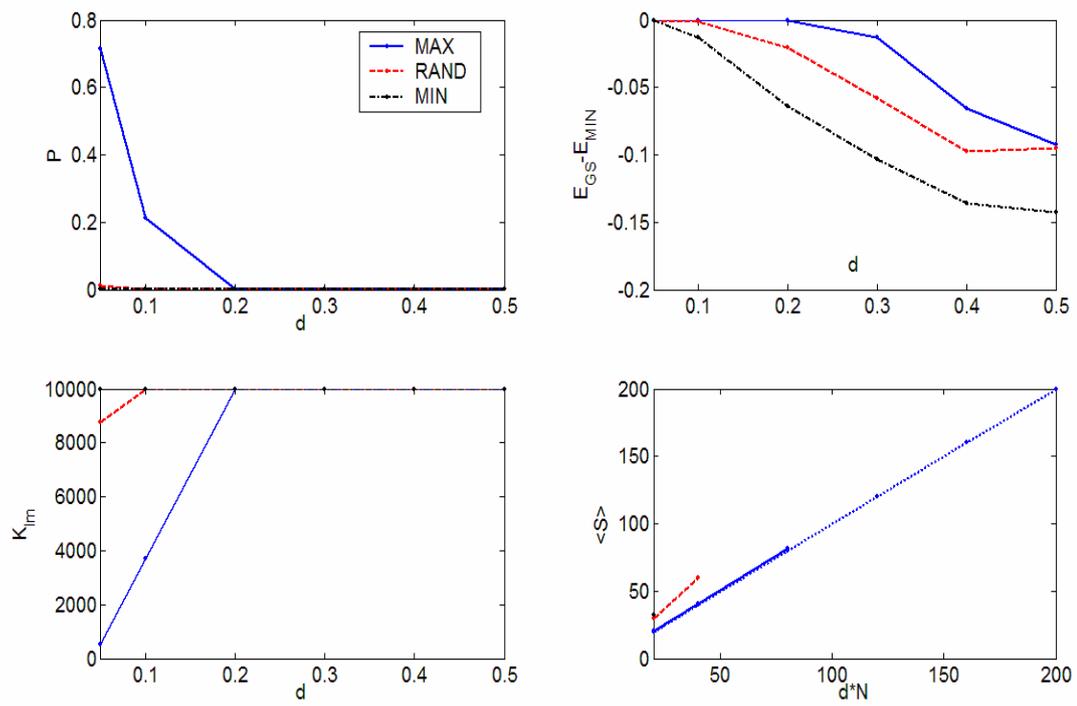

Fig.9. The same as in Fig.8 for the sin-glass matrix of the dimensionality $N=400$.

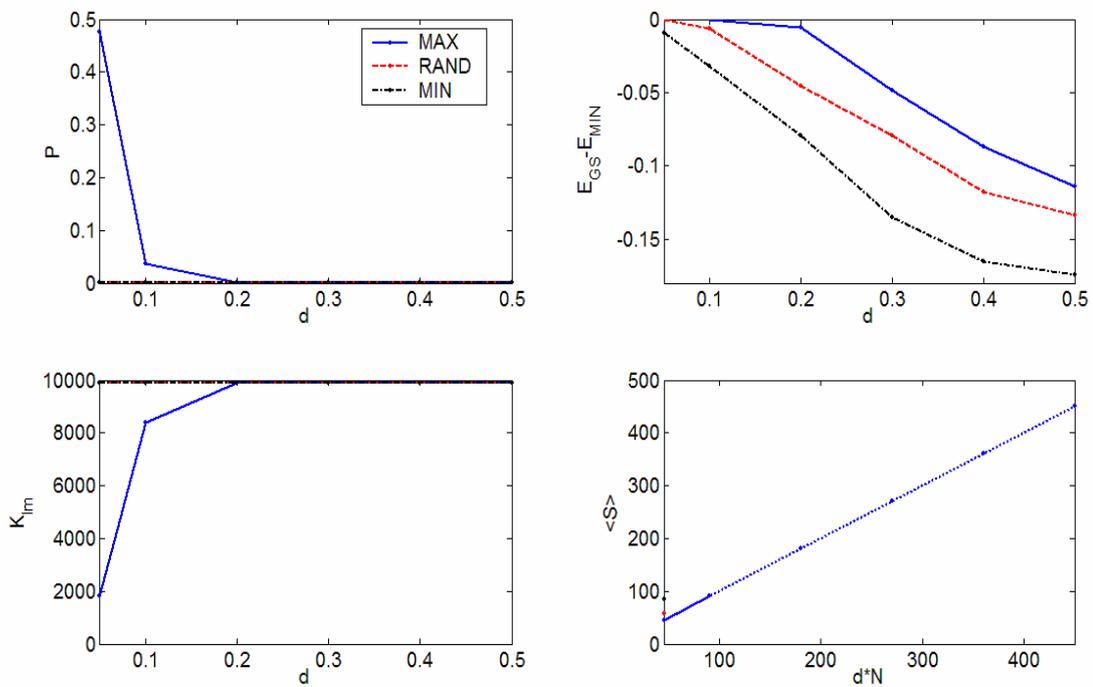

Fig.10. The same as in Fig.8 for the sin-glass matrix of the dimensionality $N=900$.

## 4. Results for the Gaussian matrices

By Gaussian matrices we shall mean symmetric matrices whose elements are independent and randomly chosen from the standard normal distribution. Six Gaussian matrices have been examined: in twos matrices of dimensionalities $N = 50$, 100 and 400. The problem we faced in the first place was how to define the ground state of the functional (1).

**1)** For Gaussian matrices of large dimensionalities there is no algorithm allowing us to define the ground state. After other authors ([11], [12], [14], [16]) we used the following approach: a great number $K \sim 10^6$ of random starts had been done, and as the ground state we chose configuration corresponding to the deepest minimum. It is desirable that the number of hits into the deepest minimum exceeds substantially $1/K$. This can guarantee that probably there is no any deeper minimum.

At Figs. 11-13 we present the results of a great number of random starts for matrices of dimensionalities $N=50$, $N=100$ and $N=400$ correspondingly (one matrix of each dimensionality). 500000 starts has been done for dimensionalities $N=50$ and $N=100$, and one million starts for dimensionality $N=400$. The graphs shown at the figures are analogous to ones shown in Fig.7. Along the abscissa axes we show the energy of local minima. At the lower panel there is the density histogram of local minima energies. At the upper panel along the axis of ordinates we show the probability to get into a local minimum. Besides that we give the number of local minima $K_{lm}$. (All calculations were done with single precision.) The random dynamics was used, and start configurations could be considered as "the points of general positions" since the mean distances between them and the ground state was equal 0.5: $<d>=0.5$.

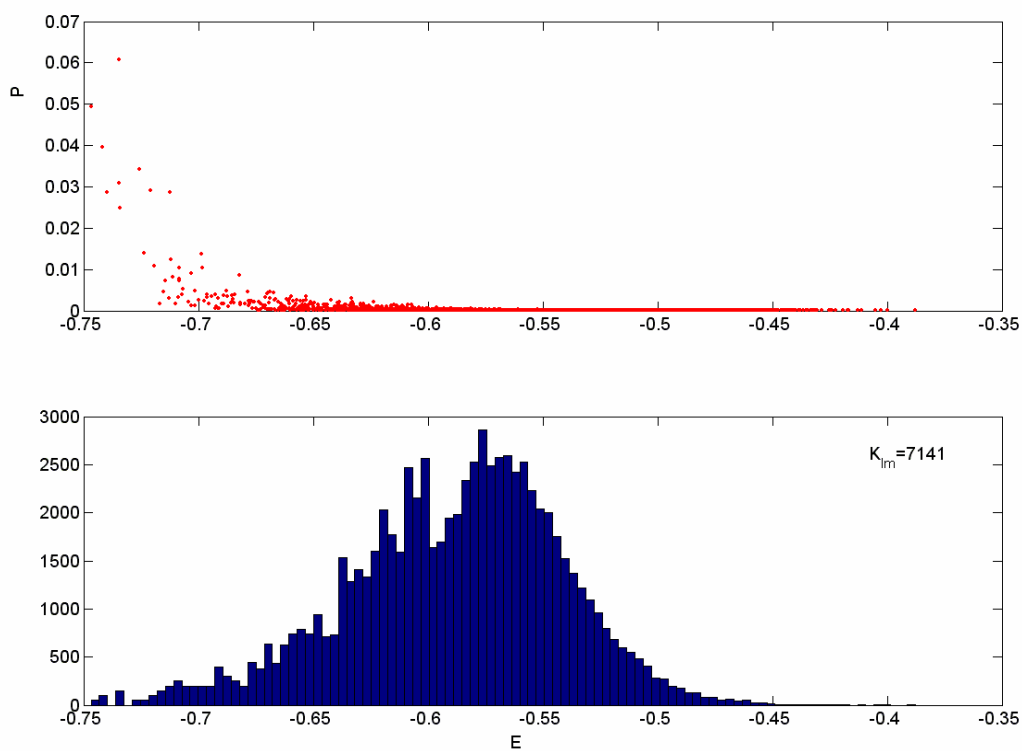

Fig.11. Results for the Gaussian matrix of the dimensionality $N=50$ for 500000 random starts.
At the lower panel is the histogram of the density of distribution of local minima. At the upper panel is the probability to get into a local minimum.

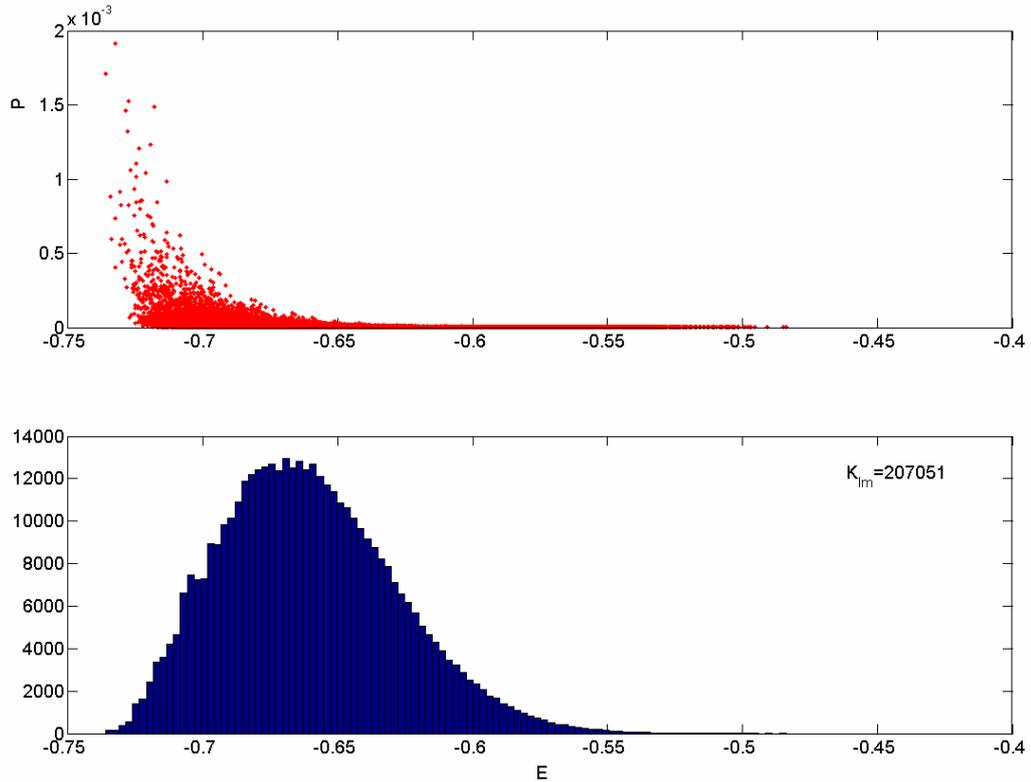

Fig.12. The same as in Fig.11 for the Gaussian matrix of the dimensionality *N*=100.

We see that for matrices of dimensionalities *N*=50 and *N*=100 the probability to get into the deepest minima substantially larger than the probability to get into less deep minima (see distributions of probabilities at upper panels of Fig.11 and Fig.12). In other words, obviously the rule of deep minima is fulfilled in these cases. Of course, the largest part of local minima corresponds to the middle of the energy interval. The hump on the histogram is evidence of that (see the lower panel of the figures). However, the probability to get into an isolated minimum is always the biggest for the deepest minima. For *N*=50 we have $P(E_{min}) \approx 0.05$, and for *N*=100 the probability $P(E_{min}) \approx 0.017$. There is good reason to believe that in these cases the configurations relating to the deepest minima are the ground states.

The situation differs for matrices of dimensionalities *N*=400. It is evident that the rule of deep minima does not fulfill in this case (see the distribution of probabilities on the upper panel of Fig.13). For million random starts the dynamical system only once gets into the deepest minimum. Can we trust in such little probability? It may be that there is a deeper minimum. To answer these questions the following was done.

According to [16], [17], for Gaussian matrices the minimal dynamics gives the best results. This means that just the minimal dynamics has to be used to find the ground state. For both matrices of dimensionality *N*=400 we did 500000 extra random starts using the minimal dynamics. For the first matrix the random and minimal dynamics provided the same configuration as the deepest minimum. However, with the aid of the random dynamics this configuration was found only once, and in the case of minimal dynamics the configuration was found 37 times. For the second matrix with the aid of minimal dynamics we found a deeper minimum than with the aid of the random dynamics. The system hits this deeper minimum 50 times. As a result for matrices of dimensionalities *N*=400 as a "ground state" we chose the configurations found with the aid of the minimal dynamics.

Describing the results for Gaussian matrices, before we proceed to the concluding part, let us note that there is a critical dimensionality $N_c$. If the dimensionality of the problem is less than $N_c$, the rule of deep minima is fulfilled. On the contrary, if the dimensionality of a matrix is greater than $N_c$, the rule of deep minima does not fulfilled. We estimated the critical value of the dimensionality as $N_c \approx 320-350$.

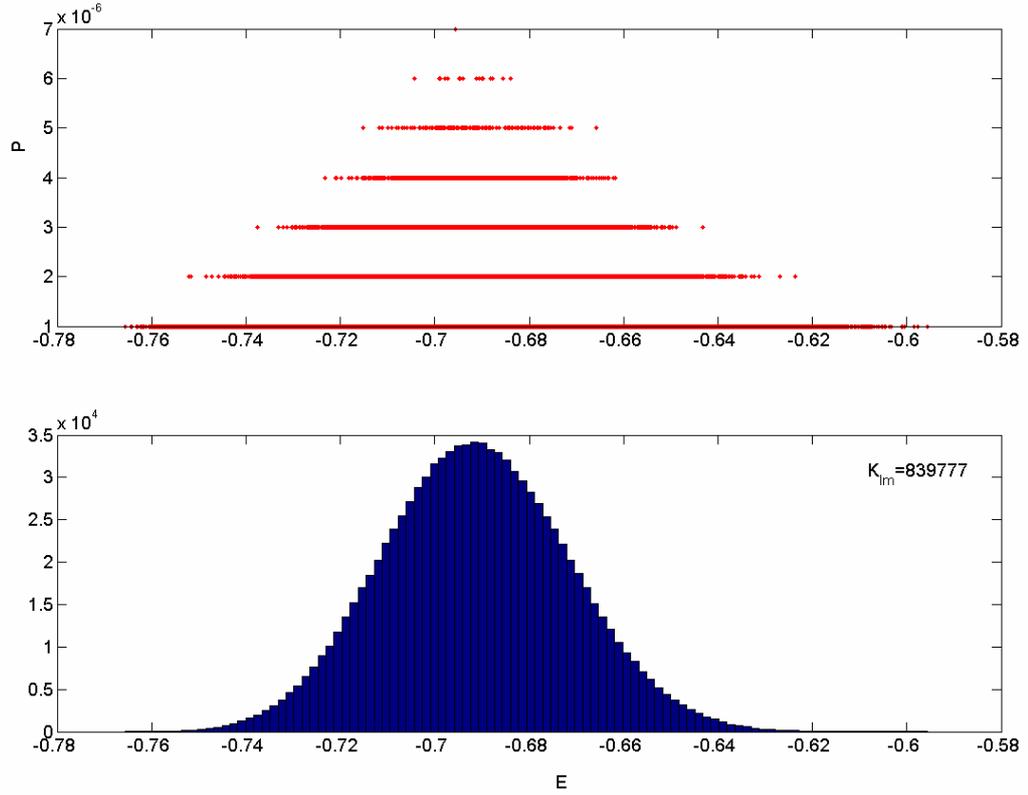

Fig.13. The same as in Fig.11 for the Gaussian matrix of the dimensionality $N=400$ ($10^6$ random starts).

**2)** The remaining part of the experiment was just the same as for spin-glass matrices. For each matrix we generated 10000 start configurations which were at the distance $d$ from the ground state. These configurations were used to start all three dynamics. The start distances $d$ were chosen the same as before. For local minima as before we fixed the same characteristics $E$, $D$ and $S$ (see the beginning of the previous Section). The results of experiments for matrices of dimensionalities $N = 50$, 100 and 400 are shown in Figs. 14-16 correspondingly. The form of presentation of the results is the same as at Figs. 8-10: the solid line corresponds to the maximal dynamics, the dashed line corresponds to the random dynamics and the dash-dotted line corresponds to the minimal dynamics. Along the abscissa axes we show the distances between the start configurations and the ground state. As before only at the right lower panels of the figures we use the Hamming distance $d \cdot N$; at the other panels along the abscissa axes we show the relative Hamming distance $d$ (3).

At the left upper panel along the ordinate axis we show the probability $P$ to find the ground state. At the panel, which is just under the abovementioned one, along ordinate axis we show the number of local minima. For start distances $d<0.4$ we see the familiar picture: the best results are given by the maximal dynamics, the worst results correspond to the minimal dynamics, and the random dynamics occupies an intermediate place. Evidently, we cannot speak about the superiority of the minimal dynamics.

Unexpectedly for start distances $d \geq 0.4$ the minimal and maximal dynamics change places. For the dimensionalities $N=50$ and $N=100$ this is clearly seen from the graphs at the left panels of Figs. 14, 15. Indeed, in these cases the probability to find the ground state is also the biggest for the minimal dynamics, as well as the number of the local minima is minimal for the same dynamics. For the dimensionality $N=400$ the superiority of the minimal dynamics is seen from the graphs at the upper right panel of Fig.16. Here along the ordinate axis we show the difference $E_{GS} - E_{\min}$ between the energy of "the ground state" and the minimal energy found after 10000 starts. We see that for the start distances $d \geq 0.4$ only the minimal dynamics allows us to find "the ground state". That cannot be done with the aid of the random or the maximal dynamics.

For all analyzed matrices we observed the superiority of the minimal dynamics for the start distances $d \geq 0.4$. Previously the superiority of the minimal dynamics was observed for start distances $d \approx 0.5$ by the authors of [16], [17]. However, let us stress that for $d < 0.4$ the minimal dynamics gives the worst results. Here the best results can be obtained with the aid of the maximal dynamics. The inversion of the minimal and the maximal dynamics for large values of $d$ is one of the characteristic distinctions of the Gaussian matrices from the spin-glass matrices.

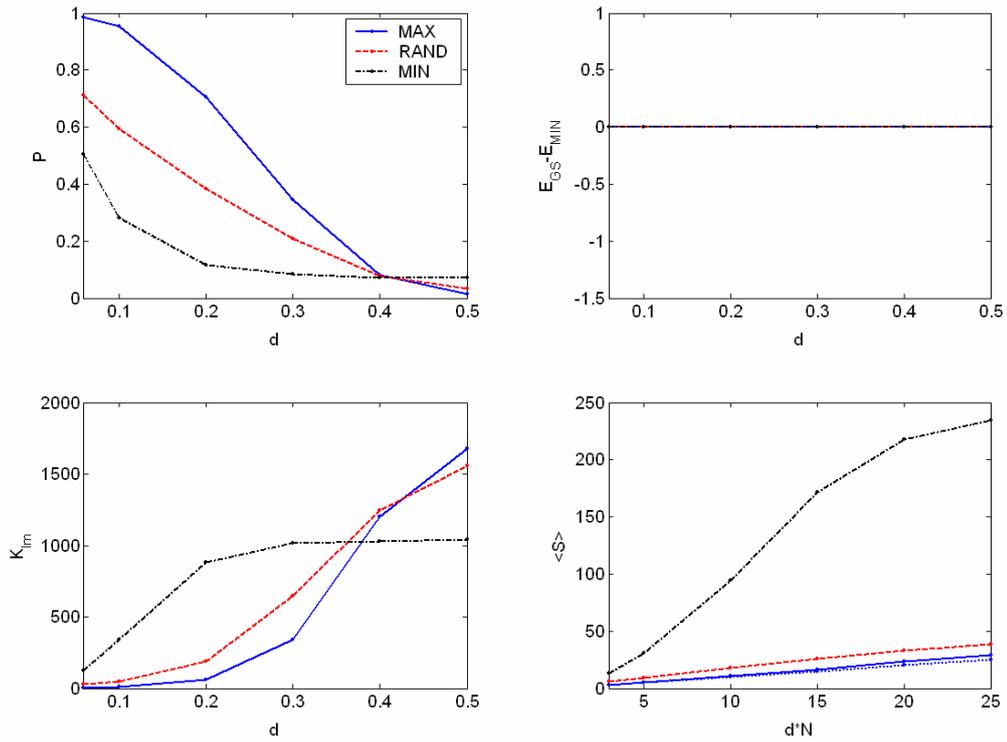

Fig.14. Results of minimization depending on start distances to the ground state for the Gaussian matrix (*N*=50). Along the abscissa axes are the Hamming distance at the right lower panel, at the other panels the relative Hamming distance *d* (3). Solid lines correspond to the maximal dynamics, dashed lines correspond to the random dynamics, and dash-dotted lines correspond to the minimal dynamics (see explanation in the body of the paper).

The second characteristic distinction is related to the behavior of the <*S*>-curve (see the right lower panels of Figs. 14-16.) The same as in the case of spin-glass matrices, the graph for <*S*> for the maximal dynamics (the solid line) practically does not differ from the straight line $<S> = d \cdot N$ (the dotted line). For the random dynamics the averaged length of the trajectory that leads the system to the ground state, somewhat bigger than for the maximal dynamics. At all figures the dashed line is situated a little higher than the solid line. And for the minimal dynamics (the dash-dotted line) the averaged length of the trajectory is at least an order larger than both aforementioned values. The overwhelming superiority of dash-dotted line is seen at the right lower panels of Figs. 14-16.

In other words, everywhere where with the aid of the maximal dynamics the ground state can be found, the system moves to it along the shortest rout: $<S> \approx d \cdot N$. This is true both for the Gaussian and spin-glass matrices. Otherwise matters stand with the minimal dynamics. For two types of matrices the averaged length of trajectory that leads the system to the ground state behaves differently. For spin-glass matrices this characteristic exceeds the start distance, but not significantly, only 1.5-2 times (see the relating graphs in Fig.8 and Fig.9). For the Gaussian matrices the averaged length of the trajectory that leads the system to the ground state is at least an order of magnitude larger than $d \cdot N$. It is evident that in this case the system moves to the ground state by an intricate roundabout way. We can say nothing more definite about the character of moving along the energy surface. The obtained results mean that for the Gaussian matrices the minimal dynamics works at least an order of magnitude slower than other dynamics. That is so since we measure the length of trajectory by the number of steps done – the number of flipped spins.

We remind that for spin-glass matrices of each dimensionality we had critical values of the start distances $d_c$, at the left of which the rule of deep minima was fulfilled, and at the right of them the rule did not fulfilled. The sign that a start distance *d* was greater than $d_c$, was an approximate (or exact) coincidence of the number of the local minima and the number of starts: $K_{lm} \approx K$. For the Gaussian matrices of not very large dimensionality $N = 50, 100$ the rule of deep minima is fulfilled for any start distances (see the distribution of probabilities to get in local minima at the upper panels of Fig.11 and Fig.12). For the Gaussian matrices of the dimensionality *N*=400 the critical value of the start distance is less than 0.4: $d_c \approx 0.35$ (see the left lower panel of Fig.16).

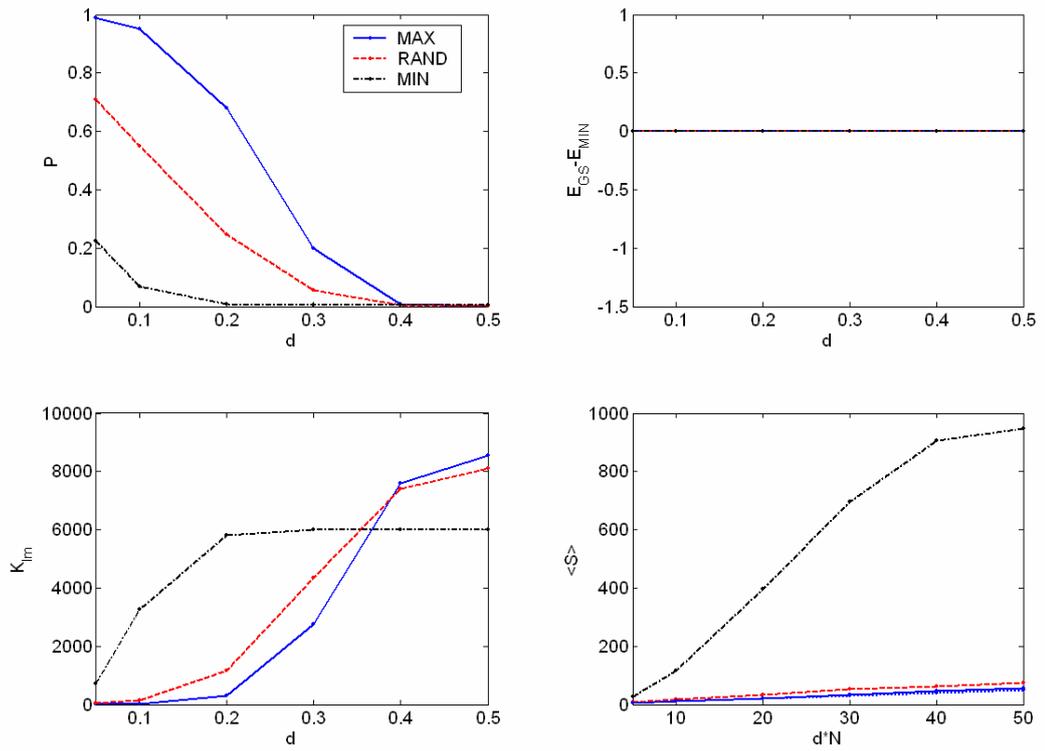

Fig.15. The same as in Fig.14 for the Gaussian matrix of the dimensionality $N$=100.

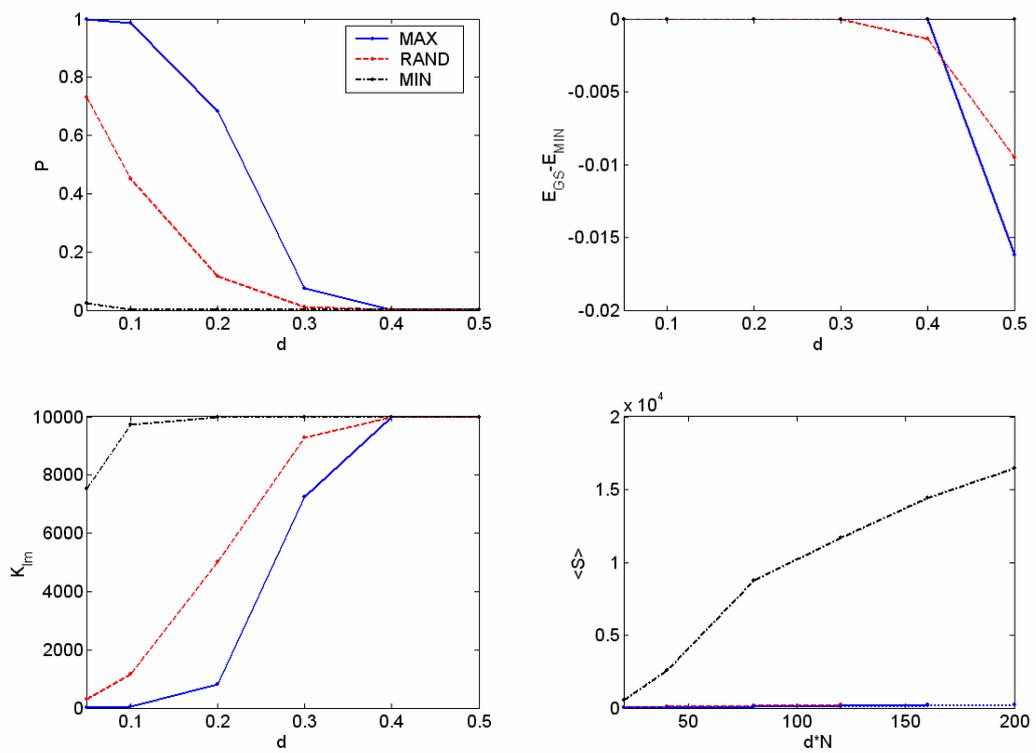

Fig.16. The same as in Fig.14 for the Gaussian matrix of the dimensionality $N$=400.

<u>Summarizing the aforesaid.</u> Till start points are away from the ground state at distances $d < 0.35$, the maximal dynamics gives the best results, the minimal dynamics gives the worst results and the random dynamics occupies an intermediate position. For $d > 0.35$ the maximal and minimal dynamics change their positions. For all values of start distances $d$ the minimal dynamics works at least an order of magnitude slower than other dynamics.

## 5. Discussion and conclusions

By means of computer simulations 3 variants of one-step minimization procedure were examined. They are the maximal, the random and the minimal dynamics. These dynamics are also known as the greedy, the sequential and the reluctant algorithms, respectively. We analyzed matrices of two types: *spin-glass* matrices (this name we use for matrices relating to the Edwards-Anderson spin-glass model) and *the Gaussian* matrices (they are relating to the Sherrington-Kirkpatrick spin-glass model). The dimensionalities of the matrices were from the interval $N \in [50,1000]$. The main conclusion is that as a rule the maximal dynamics demonstrates notably better results than other dynamics. For spin-glass matrices the superiority of the maximal dynamics takes place for all values of start distances. For the Gaussian matrices ($N < 500$) the superiority of the maximal dynamics takes place only for start distances $d < 0.35$; for larger values of start distances, $d \in [0.4, 0.5]$, the minimal dynamics allows one to find deeper minima (see also [16], [17]). Note, for the Gaussian matrices the minimal dynamics works much slower (1-2 orders of magnitude) than the maximal and random dynamics.

Two most significant distinctions in the results obtained for these two types of matrices are related with the minimal dynamics. For the Gaussian matrices and large values of start distances just the minimal dynamics allows us to obtain the best results. Against almost total superiority of the maximal dynamics, this "peak" of efficiency of the minimal dynamics appears rather unexpected. Also unexpected is the fact that in the case of the Gaussian matrices the minimal dynamics leads to very long and intricate trajectories along which the system gets to the ground state. The lengths of these trajectories more than an order of magnitude larger than the distance between a start point and the ground state. It can be assumed that the reason of abovementioned distinctions is differences of energy landscapes relating to these two types of matrices. However, we do not know what these differences are. The general impression is that for the Gaussian matrices the energy surface is "smoother". Its "sides" are less "speckled" by local minima than in the case of spin-glass matrices. This is just the reason why for the Gaussian matrices one can find the ground state almost always, and for spin-glass matrices this can be done only when the start points are not too far from the ground state.

For the matrices of both types the rule of deep minima is fulfilled, namely: the probability to get into deeper minima is greater than the probability to get into less deep ones (see Introduction and [19], [20]). In the same time not always this rule is fulfilled. For spin-glass matrices of all analyzed dimensionalities it is failed when start distances are larger than some critical values $d_c(N)$. In this case the probability to get into the ground state is negligibly small. For the Gaussian matrices of dimensionalities $N \leq 300$ the rule of deep minima is fulfilled for all start distances. In the same time, for the matrices of the dimensionality $N=400$ this rule failed, when start distances exceed the critical value $d_c \approx 0.35$.

The superiority of the maximal dynamics is the greater, the less the distance of a start point to the ground state. Just in such situation the maximal dynamics is most effective. Can we choose start points that are near the ground state, if we do not know the ground state itself? Authors of the works [23], [24] shown that it is possible. They introduced and analyzed the so called *EIGEN-configurations*. These are configurations that are the nearest to the eigenvectors of the connection matrix. Some of EIGEN-configurations possess useful properties. In particular, the *greatest* EIGEN-configurations, which are close to the greatest eigenvectors. For example, 10-20 maximal EIGEN-configurations correspond to very low energies. These start configurations are already at the energy level of the local minima; this is true for matrices of any types and any dimensionalities. It has been found out (see [25], [26]) that starts from these configurations led the system to the deepest minima. In the present work it is shown experimentally that among the maximal EIGEN-configurations there are certainly 2-3 configurations that are situated at the distance $d \approx 0.25 - 0.3$ from the ground state. As a rule they are the EIGEN-configurations that are the nearest to 2-3 largest eigenvectors. This result is true for both types of the examined matrices and all dimensionalities $N$. We hope to use this result for effective constructing of a whole set of random configurations that are situated at the distances $d \approx 0.25 - 0.3$ from the ground state. May be these configurations would be good start points for founding the ground state.

**Acknowledgements**

The author is grateful to D.V. Vylegzhanin who took part at initial stage of this work. Also the author is grateful to participants of the seminar of Center of Optical-Neural Technologies of Scientific Research Institute for System Analysis of RAS where this work was presented. The work was done in the framework of the program "Information Technologies and an Analysis of Complex Systems" under financial support of Russian Basic Research Foundation Grant No. 09-07-00159.